\documentclass[12pt, nofootinbib]{article}
\renewcommand{\title}[1]{\vbox{\center\LARGE{#1}}\vspace{5mm}}
\renewcommand{\author}[1]{\vbox{\center#1}\vspace{5mm}}
\newcommand{\address}[1]{\vbox{\center\em#1}}

\renewcommand{\date}[1]{\vbox{\center#1}}

\usepackage{amssymb}
\usepackage{amsmath,bm}
\usepackage{amssymb}
\usepackage{graphicx}
\usepackage{amsfonts}         
\usepackage{fancybox}

\usepackage{enumitem}

\usepackage{slashed}

\usepackage[font=small,labelfont=bf]{caption}
\DeclareCaptionFont{tiny}{\tiny}
\usepackage{yfonts} 
\usepackage{epstopdf}
\usepackage[usenames,dvipsnames]{xcolor}%http://en.wikibooks.org/wiki/LaTeX/Colors
\usepackage[pdftex, bookmarks={false}, pdfauthor={John McGreevy}, pdftitle={This is a secret message!}]{hyperref}
\hypersetup{colorlinks=true, linkcolor=BrickRed, citecolor=Violet, filecolor=OliveGreen, urlcolor=RoyalBlue, filebordercolor={.8 .8 1}, urlbordercolor={.8 .8 0}}%http://en.wikibooks.org/wiki/LaTeX/Hyperlinks
\usepackage{soul}%strike through text \st{} .  not in math mode.
\setstcolor{Red}

\usepackage{wrapfig}
\definecolor{darkgreen}{rgb}{0,0.4,0}
\definecolor{darkred}{rgb}{0.4,0,0}
\definecolor{darkblue}{rgb}{0,0,0.4}
\definecolor{lightblue}{rgb}{.6,.6,0.9}

\newcommand{\cob}{\color{black}}

\definecolor{uglybrown}{rgb}{0.8,  0.7,  0.5}

\definecolor{palatinatepurple}{rgb}{0.41, 0.16, 0.38}
\definecolor{celebrationcolor}{rgb}{0.75,  0.0,  0.9}

 \usepackage{mdframed}

\usepackage{framed}
\definecolor{shadecolor}{rgb}{0.90,0.90,0.90}

\usepackage{wasysym}
\def\subsubsubsection#1{{\bf #1}}

\def\breakS{\vfill\eject}
\numberwithin{equation}{section}

\renewcommand{\theequation}{\arabic{section}.\arabic{equation}}

\def\nd{{ \vphantom{\dagger}}}

\input epsf
\newcommand{\vev}[1]{\langle #1 \rangle}

\newlength{\extraspace}
\newlength{\extraspaces}
\def\be{\begin{equation}}
\def\ee{\end{equation}}

\newcommand{\bea}{\begin{eqnarray}}
\newcommand{\eea}{\end{eqnarray}}

\def\eps{\epsilon}
\def\half{{1\over 2}}

\def\bra#1{{\langle}#1|}
\def\ket#1{|#1\rangle}

\def\vev#1{\langle{#1}\rangle}

\def\CB{{\cal B}}

\def\CF{{\cal F}}

\def\CM{{\cal M}}
\def\CN{{\cal N}}
\def\CT{{\cal T}}

\def\CV{{\cal V}}

\def\CX{{\cal X}}%AEL

\def\CZ{{\cal Z}}

\def\sgn{{\rm sgn\ }}

\def\II{\relax{I\kern-.10em I}}

\def\IZ{\mathbb{Z}}
\def\IB{\relax{\rm I\kern-.18em B}}

\def\ID{\relax{\rm I\kern-.18em D}}
\def\IE{\relax{\rm I\kern-.18em E}}
\def\IF{\relax{\rm I\kern-.18em F}}
\def\IG{\relax\hbox{$\inbar\kern-.3em{\rm G}$}}
\def\IGa{\relax\hbox{${\rm I}\kern-.18em\Gamma$}}
\def\IH{\relax{\rm I\kern-.18em H}}
\def\II{\relax{\rm I\kern-.18em I}}
\def\IK{\relax{\rm I\kern-.18em K}}
\def\inbar{\,\vrule height1.5ex width.4pt depth0pt}

\def\gs{g_s}
\def\lp10{\ell_p^{10}}
\def\lp11{\ell_p^{11}}
\def\R11{R_{11}}

\def\frac#1#2{{#1 \over #2}}

\def\up{\uparrow}
\def\down{\downarrow}

\newdimen\tableauside\tableauside=1.0ex
\newdimen\tableaurule\tableaurule=0.4pt
\newdimen\tableaustep
\def\phantomhrule#1{\hbox{\vbox to0pt{\hrule height\tableaurule width#1\vss}}}
\def\phantomvrule#1{\vbox{\hbox to0pt{\vrule width\tableaurule height#1\hss}}}
\def\sqr{\vbox{%
  \phantomhrule\tableaustep
  \hbox{\phantomvrule\tableaustep\kern\tableaustep\phantomvrule\tableaustep}%
  \hbox{\vbox{\phantomhrule\tableauside}\kern-\tableaurule}}}
\def\squares#1{\hbox{\count0=#1\noindent\loop\sqr
  \advance\count0 by-1 \ifnum\count0>0\repeat}}
\def\tableau#1{\vcenter{\offinterlineskip
  \tableaustep=\tableauside\advance\tableaustep by-\tableaurule
  \kern\normallineskip\hbox
    {\kern\normallineskip\vbox
      {\gettableau#1 0 }%
     \kern\normallineskip\kern\tableaurule}%
  \kern\normallineskip\kern\tableaurule}}
\def\gettableau#1 {\ifnum#1=0\let\next=\null\else
  \squares{#1}\let\next=\gettableau\fi\next}

 \def\eqnn#1{\xdef #1{(\secsym\the\meqno)}\writedef{#1\leftbracket#1}%
 \global\advance\meqno by1\wrlabeL#1}
 \def\eqna#1{\xdef #1##1{\hbox{$(\secsym\the\meqno##1)$}}
 \writedef{#1\numbersign1\leftbracket#1{\numbersign1}}%
 \global\advance\meqno by1\wrlabeL{#1$\{\}$}}
 \def\eqn#1#2{\xdef #1{(\secsym\the\meqno)}\writedef{#1\leftbracket#1}%
 \global\advance\meqno by1$$#2\eqno#1\eqlabeL#1$$}

\global\newcount\itemno \global\itemno=0

\def\itemaut#1{\global\advance\itemno by1\noindent\item{\the\itemno.}#1}

\def\({\left(}
\def\){\right)}

\def\ii{{\bf i}}

\def\HH{{\bf H}}

\def\UU{{\bf U}}

 \def\XX{{\bf X}}
 \def\ZZ{{\bf Z}}

\def\lsim{\mathrel{\mathstrut\smash{\ooalign{\raise2.5pt\hbox{$<$}\cr\lower2.5pt\hbox{$\sim$}}}}}
\def\gsim{\mathrel{\mathstrut\smash{\ooalign{\raise2.5pt\hbox{$>$}\cr\lower2.5pt\hbox{$\sim$}}}}}

\def\overleftrightarrow#1{\vbox{\ialign{##\crcr
     $\leftrightarrow$\crcr\noalign{\kern-0pt\nointerlineskip}
     $\hfil\displaystyle{#1}\hfil$\crcr}}}
     
     \def\overleftarrow#1{\vbox{\ialign{##\crcr
     $\leftarrow$\crcr\noalign{\kern-0pt\nointerlineskip}
     $\hfil\displaystyle{#1}\hfil$\crcr}}}

\def\eg{{\it e.g.}}
\def\ie{{\it i.e.}}

\def\gG{\textsf{G}}

\def\gSU{\textsf{SU}}

\def\gS{\textsf{S}}

\def\gs{\text{gs}}

\usepackage[yyyymmdd,hhmmss]{datetime}
\newif{\ifeq}           % defines a new condition @eq tested by the conditional \ifeq
\eqtrue                 % if uncommented, declares @eq to be true
\newcounter{lecturecounter}
\usepackage{wasysym}

\begin{document}
%\maketitle
%\section{}
%\subsection{}

\def\sZ{\ZZ}
\def\sX{\XX}
\def\CX{\mathfrak{X}}

%\title{Brief Article}
%\author{The Author}

%\date{}							% Activate to display a given date or no date
\begin{titlepage}

%\title{Exactly solvable models of ``decorated domain walls"}
% I do not like this title.
\title{Worldsheet matter for electric flux strings \\
{\footnotesize or }\\
%\title{
Exactly solvable models of spin liquids with spinons, and of 3d topological paramagnets\\
{\footnotesize or } \\
SNAKE MONSTERS!
}

\author{
Daniel Ben-Zion, Diptarka Das, John McGreevy}

\address{Department of Physics, University of California at San Diego, La Jolla, CA 92093, USA}

\begin{abstract}
We develop a scheme to make exactly solvable gauge theories
whose electric flux lines host (1+1)-dimensional topological phases.
We use this exact `decorated-string-net' framework
to construct several classes of interesting models.
In particular, we construct an exactly solvable model 
of a quantum spin liquid whose (gapped) 
elementary excitations form doublets under an internal symmetry,
and hence may be regarded as spin-carrying spinons.
The model may be formulated, and is solvable, in any number of dimensions,
on any bipartite graph.
%(If you read a little further we will come clean about the size of the symmetry group.)
Another example, in any dimension, 
has $\IZ_2$ topological order and 
anyons which are Kramers' doublets of time reversal symmetry.
Further, we make exactly solvable models of 
%several
3d topological paramagnets. 

\end{abstract}

\vfill
\today

\end{titlepage}
\vfill\eject
%blank.
%\vfill\eject
\setcounter{tocdepth}{2}    % don't show subsubsections in toc
\parskip = 0.02in

%{{font=footnotesize}
\tableofcontents

\parskip = 0.1in

\vfill\eject

\section{Context}

\subsubsubsection{Worldsheet matter for electric flux strings.} The idea that worldsheets of strings may have 
dynamical degrees of freedom living upon them
(in addition to the fields which encode their embedding in space(time)) 
is crucial for fundamental string theory \cite{Polchinski:1998rr}. 
This possibility is realized in other contexts as well,
such as in
%For example: Polyakov's derivation of 
the worldvolume theory of 
domain wall strings 
in the 2d Ising model \cite{Polyakov:1988qz}.

In this paper we are going 
to show how to glue (1+1)-dimensional topological states to 
the electric flux lines of a gauge theory,
in an exactly solvable way.
The signature of a nontrivial 1d topological state 
is some degeneracy at the edge of an 
open chain, generally 
representing projectively a symmetry of the system.
Since the ends of electric flux lines 
are electric charges, our construction 
provides a machine for 
imbuing the charges of a gauge theory 
with nontrivial symmetry properties.

The information we need to accomplish this goal is just
a certain ``circuit construction" of the 1d topological state, 
\ie~a collection of unitary operators associated to the links which create and destroy
the desired 1d state out of a background bath of product states.

Previously, certain models of `decorated domain walls' have been constructed which are frustration free; only the ground state is known \cite{2014NatCo...5E3507C, 2015arXiv150600592P}. Here, we systematically generalize this idea to produce a variety of interesting decorated string net models whose entire spectrum is known.  
In particular, we can guarantee that the spectrum is gapped.
We will occasionally refer to these models as snake monsters. The name is motivated by the idea that we are imparting dynamics to these one-dimensional creatures
with all the action at the ends. We focus most of our attention on two and three spatial dimensions, although extensions to higher dimensions are readily possible.

\subsubsubsection{What are these models for?}
It has been known for some time that the possibilities for phases of quantum matter extend 
far beyond Landau's symmetry-breaking paradigm \cite{wen04}. 
Two groundstates can preserve the same symmetry yet nonetheless belong to distinct phases.
Even in the absence of symmetry, different patterns of long range entanglement can lead to distinct types of topological order.

By now there exist several partial classification schemes for topological phases;
it is not clear that these schemes are complete.
Furthermore, these methods are quite formal and do not necessarily lead to an intuitive understanding of the physics. 
Therefore, it is worthwhile to search for tractable examples which realize interesting phases.

\subsubsubsection{Herding snakes.} 
In section \ref{sec:snake-def} we introduce
the scheme in somewhat general terms.
Since this construction is rather versatile and has already led us to a 
%perhaps bewildering array 
variety of examples, 
we first provide some organizing discussion and explain how they differ in physics and in 
technical aspects.
The models we discuss can be organized along several axes:
\begin{enumerate}
\item {\it Do they have topological order, and of what nature?}

The models discussed in sections \ref{sec:snake-def}-\ref{sec:cluster-snake} have (abelian) topological order.
These are therefore solvable representatives of symmetry-enriched topological (SET) phases.
Previous examples 
of solvable models (and indeed partial classifications) of such phases have appeared in \cite{2013PhRvB..87j4406E, 2013PhRvB..87o5115M},
but our approach is quite different.
In this context, the novelty of our construction is its simplicity and flexibility.

In section \ref{sec:3d-paramagnets} we extend a
construction \cite{wns2015} 
of 3d models made from fluctuating strings
{\it without} topological order.
We use this to make solvable models of some topological paramagnets,
and discuss their nontriviality as symmetry-protected topological (SPT) states\footnote{For reviews
of SPT and SET physics, see \cite{Turner:2013kp,Senthil:2014ooa}.  Exactly solvable models
have been constructed for some examples in \cite{2012Levin-Gu,2014arXiv1403.6491C,
2013PhRvX...3d1016F, 2015arXiv150803468Y}.}.
The phases represented by this framework go beyond
the group cohomology classification of \cite{PhysRevB.87.155114}.

\item {\it What symmetry protects their nontriviality?}

In section \ref{sec:1dreview} we provide a detailed review of 
cluster states, 
an example of a 1d SPT.
These enjoy a solvable Hamiltonian 
which is 
protected from triviality either by  time-reversal symmetry
or (on a bi-partite graph) by a unitary $ \IZ_2\times \IZ_2$ symmetry.
%In \S\ref{sec:AKLT} we introduce a set of time reversal symmetric models
%with an additional $ \gU(1)$ symmetry.

\item {\it Is it a model of bosons/spins or one with microscopic fermions?}

An example of the former may be obtained from an example of the latter 
by  gauging the fermion parity symmetry.
(A recent discussion of this connection appears in \cite{2014arXiv1404.6256Y}.)

%Through \S\ref{sec:z-per-site} 
% In the body of the paper,
% all our examples
% are models of bosons, with no gauge-neutral fermionic excitations.
% In \S\ref{sec:majorana-subsection} we 
% construct a model with microscopic fermions
% enjoying time-reversal symmetry.
% %{\cor THIS SEEMS NOT TO BE THE CASE:} We proceed in \S\ref{sec:Z4snakes} 
% We leave it for the future
% to gauge the fermion number symmetry
% and thereby make a boson SET model.
In this paper, all of our examples are models of bosonic SPTs. 
One can construct fermionic SPTs along similar lines, but we leave that for the future. 
\end{enumerate}

\section{Snake monster}

\label{sec:snake-def}

We start with an exactly solvable lattice gauge theory. We introduce additional degrees of freedom coupled to the gauge fields in a nontrivial way to imbue the gauge theory with further symmetry properties. As a result of this procedure, anyons carry projective representations of the symmetry and we show that these models realize distinct phases. The simplest context in which to introduce our construction is $\IZ_2$ lattice gauge theory.

\noindent{\bf Toric code review.}
To establish notation, recall the 
toric code \cite{Kitaev:1997wr}, 
a system of qbits on the links of a graph 
which emerges $\IZ_2$ gauge theory.
The toric code is governed by the following hamiltonian: 
$$ \HH_\text{TC} = - \sum_p B_p -  \sum_j A_j ~$$
where $j$ runs over sites of the graph,
and $p$ runs over the faces\footnote{Actually we are using a bit more structure than just a graph.
The required information is a simplicial complex: 
a list of $p$-dimensional subspaces $ \Omega^p$ and a 
boundary map $\partial: \Omega^p \to \Omega^{p-1}$ which
says who is the boundary of whom.  For $\IZ_2$,
the orientations do not matter.}
and
$$ B_p \equiv \prod_{l \in \partial p} \sigma^x_l, ~~~~A_j \equiv \prod_{l \in v(j)}  \sigma^z_l . $$
(Here $\partial p$ denotes the collection of links 
in the boundary of the plaquette $p$ and $v(j)$ is the `vicinity operator'
which gives the collection of links whose boundary contains the site $j$.)
These operators commute and can be simultaneously diagonalized. A useful description of a state is to imagine a link as occupied by a string if $\sigma^z_l = -1$ and unoccupied if $\sigma^z_l = +1$. Then satisfying the `star terms' ($A_i\ket{\psi} = \ket{\psi} \forall i$) means that strings do not end; arbitrary superpositions of closed strings are the ground states. 

%These operators commute.  
%Diagonalize $ A_s$ first: arbitrary superpositions of closed strings of edges %where $ \sigma^z_l=-1$ 
%are the groundstates.

The closed-string states of the link variables are
$$ \ket{C} = \prod_{l \in C} \sigma^x_l \otimes_{l} \ket{\sigma^z = 1 } ~~$$
where $C$ denotes a collection of occupied links.
Their degeneracy under $ \sum_j A_j$ is split by the action of $B_p$, which acts as a kinetic term for the strings: 
$$ B_p \ket{C} = \ket{C + \partial p} $$
The eigenvalue condition $B_p = 1$ 
then demands that the groundstate wavefunctions $\Psi(C) \equiv \vev{C | \text{groundstate}} $ 
have equal values for cycles $C$ and $C' = C + \partial p$.
This is the equivalence relation defining the $1$st homology of the simplicial complex:
distinct, linearly-independent groundstates are the labelled by
%$$-homology classes of $\DDD$.
homology classes with coefficients in $\IZ_2$.
%, $H^p(\DDD, \IZ_N)$.
On a simply connected space, there is a unique groundstate
$$ \ket{\text{gs}} = {1\over\sqrt{\CN_C}} \sum_\text{closed~string~collections, $C$} \ket{C}$$
where $ \CN_C $ is the number of closed string configurations. 

%$\CN_C$ is a normalization factor.

\noindent{\bf Circuit description of 1d states.} Suppose we are given a {\it circuit construction} 
of a nontrivial state of a chain $c$ of quantum spins: 
\be \label{eq:chain-state-c} \ket{c}  = U \otimes_j  \ket{\rightarrow_j} .\ee
The operator  
$ U \equiv \prod\limits_{l} u_l $ is a product of local unitaries acting on the links which creates the state $\ket c$ from a reference product state.
We consider here the case where the 1d state is 
classified by $\IZ_2^n$ for some $n$.
We assume the following properties of the link unitaries:
\begin{itemize}
\item $u_l^2 =1.$ 
%$\( \prod_{l\in C} u_l \)^2 = 1 $ if $C$ is a closed chain.   
\item  $ [ u_l, u_{l'}] = 0 $.
\item 
When we say that the 1d state is nontrivial, we mean that 
it cannot be turned into a product state by
acting with any finite-depth circuit which respects some given symmetry operation.
This in turn imposes that individual link operators $u_l$ fail to commute with the symmetry operation.
\end{itemize}
 We give an example of a collection of link unitaries satisfying our demands in \S\ref{sec:1dreview}.

\noindent{\bf Snake monster.} 
Finally, consider a system with both $\IZ_2$ link variables (whose Pauli operators we call $ \sigma^x, \sigma^z$ as above)
and site variables (whose Hilbert space we do not specify yet, but on which the $ u_l$ act).
%(whose Pauli operators are called $X, Z$).
Let $$ {\cal B}_p\equiv B_p \prod_{l \in \partial p} u_l $$
and $$ h_j \equiv \prod_{ l \in v(j) } u_l^{s_l}  h_j^0  \prod_{ l \in v(j) } u_l^{-s_l} . $$
Here $s_l \equiv \tfrac{1}{2} ( 1 - \sigma^z_l)$ counts the number of electric flux lines on the link $l$ which 
in a $ \IZ_2$ gauge theory
takes values $\{0,1\}$.
The general snake monster hamiltonian is:
\be\label{eq:general-snake-hamiltonian}
 \HH = - \sum_p {\cal B}_p -  \sum_j A_j + \sum_j  h_j P_j \ee
where 
$ P_j \equiv \tfrac{1}{2} \( 1 + A_j \) $
is the projector onto locally closed strings at site $j$.

%\label{sec:hs-does-not-matter} 
The Hamiltonian is a sum of commuting terms and is therefore exactly solvable,
as long as $h_j^0$ is exactly solvable.
This statement is otherwise independent of the form of $h_j^0$.
\footnote{ To see this explicitly,
consider a plaquette sharing two links $ l_1, l_2$ with a site term $H_s$. Then, ignoring terms which commute trivially, 
% Let 
% $$ {\cal B}_p\equiv 
% B_p \prod_{l \in \partial p} u_l = \sigma^x_{l_1} \sigma^x_{l_2} ... u_{l_1} u_{l_2} ... $$
% and
% $$ H_s = \prod_{ l \in v(s) } u_l^{s_l}  h_s  \prod_{ l \in v(s) } u_l^{-s_l}  
% = u_{l_1}^{s_{l_1}} u_{l_2}^{s_{l_2}}  ...  h_s...  u_{l_1}^{-s_{l_1}} u_{l_2}^{-s_{l_2}}  $$
% (... means factors which commute trivially).
%thinking a little bit about the algebra question we set up
%i think it works,
\bea \label{eq:hs-does-not-matter} 
{\cal B}_p h_j 
&=& u_{l_1}^{(s_{l_1}-1) } u_{l_2}^{ ( s_{l_2} - 1 ) }   u_{l_1} u_{l_2} ..  h_j
.. u_{l_1}^{-s_{l_1}+1 } u_{l_2}^{-s_{l_2} + 1  }     \sigma^x_{l_1} \sigma^x_{l_2} .. 
\cr &=&( u_{l_1} u_{l_2} )^{1-1} 
\underbrace{  u_{l_1}^{(s_{l_1}) } u_{l_2}^{ ( s_{l_2}  ) } ..  h_j
.. u_{l_1}^{-s_{l_1} } u_{l_2}^{-s_{l_2}   }  }_{= h_j} 
\underbrace{ u_{l_1} u_{l_2} ..      \sigma^x_{l_1} \sigma^x_{l_2} .. }_{= {\cal B}_p}
\cr &=& h_j {\cal B}_p .
\eea
}
We choose $h_j^0$ such that its unique ground state is a product state.  

\noindent {\bf The groundstate.}
Let $U_{C} = \prod\limits_{l \in C}u_l$ denote the product of link unitaries over links in a collection of strings $C$. The groundstate of $\HH$ on a simply connected space is: 
\bea
\ket{\text{gs}} &=& {1\over \sqrt{\CN_C}} \sum_{\text{closed~string~collections}, C} 
\ket{C}
\otimes
 U_C \otimes_i \ket{\rightarrow_i} 
\cr 
&=& 
{1\over \sqrt{\CN_C}} \sum_C \( \prod_{l \in C} \sigma^x_l u_l \) 
\(  \otimes_{l} \ket{\sigma^z_l = 1 } \otimes_i \ket{\rightarrow_i} \) 
\cr 
&=& 
{1\over \sqrt{\CN_C}} \sum_C 
\prod_{p \in R| \partial R = C}  {\cal B}_p 
\Big(  \otimes_{l} \ket{\sigma^z_l = 1 } \otimes_i \ket{\rightarrow_i} \Big) ~.
\eea

% \bea
% \ket{\text{gs}} &=& {1\over \sqrt{\CN_C}} \sum_{\text{closed~string~collections}, C} 
% \ket{C}
% \otimes
% \prod_{\text{strings}, c \in C} S_c \otimes_i \ket{\rightarrow_i} 
% \cr 
% &=& 
% {1\over \sqrt{\CN_C}} \sum_C \( \prod_{l \in C} \sigma^x_l \prod_{\text{strings}, c \in C} S_c   \) 
% \(  \otimes_{l} \ket{\sigma^z_l = 1 } \otimes_i \ket{\rightarrow_i} \) 
% \cr 
% &=& 
% {1\over \sqrt{\CN_C}} \sum_C 
% \prod_{p \in R| \partial R = C}  {\cal B}_p 
% \(  \otimes_{l} \ket{\sigma^z_l = 1 } \otimes_i \ket{\rightarrow_i} \) ~.
% \eea

To summarize the preceding construction,
the groundstate of the site hamiltonian 
$h_s$ puts the site variables along electric flux lines of the gauge theory into the state $\ket c$ 
in \eqref{eq:chain-state-c},
while putting the rest into a product state. The plaquette and site terms commute because $\CB_p$ simultaneously moves the flux line and the path along which $\ket c$ is laid. 

Notice that we have not had to specify the number of spatial dimensions. 
In two dimensions, our construction is similar to decorated-domain-wall models, in that it involves the decoration of fluctuating closed strings. In three dimensions, domain walls are two dimensional surfaces rather than strings, so this analogy fails. Furthermore, 
%the possibility of non-contractible strings is nonexistent in a model of 
domain walls are by definition contractible, whereas 
non-contractible string configurations are allowed and present in the cases we consider. Our scheme therefore naturally extends the idea of decorated domain walls to realize decorated string nets. 

\subsection{Generalization to other quantum double models}
\label{sec:2.1}
An extension to $\IZ_n$ gauge theory will occasionally be useful.
Now we must choose an orientation for each element of our simplicial complex,
and the boundary map keeps track of signs.
Place an $n$-state hilbert space on each link,
with clock operators $ \sigma^x \sigma^z = \omega^{-1}\sigma^z \sigma^x, \omega \equiv e^{ 2 \pi \ii/n}$.
\be\label{eq:ZN-snake-hamiltonian} \HH = - \sum_p \(  {\cal B}_p + h.c.\) -  \sum_j \(  A_j + h.c. \)  + \sum_j  h_j P_j\ee
Here $A_j  = \prod_{l \in v(j) }  \sigma^z_l $
-- in this product, the links are taken to point away from the site $j$ --
and  $ {\cal B}_p = B_p U_{\partial p} $ 
with
$ B_p = \prod_{l \in \partial p } \sigma^x_l  $ 
and $U_{\partial  p} = \prod_{l \in \partial_p } u_l $.
The link unitaries $u_l$ act on site variables at the ends of the links 
and we assume that\footnote{
The construction
could be generalized for unitaries which only represent $ \IZ_n$ on closed chains:
($ \left( \prod_{l\in C} u_l \right)^n = 1 $ if $C$ is a closed chain).
This condition means that $u_{jk}^n = v_j w_k$, 
\ie, that $u$ is a $\IZ_n$ operation modulo
an on-site basis change.  
%{\cobl [When can such a $u$ be written
%in terms of a $u$ satisfying our conditions? Connection to group cohomology.  Refs.  ] }
}
\begin{itemize}
%\item $ \( \prod_{l\in C} u_l \)^n = 1 $ if $C$ is a closed chain.   
\item  $ [ u_l, u_{l'}] = 0 $.
\item $u_l^n=1$.
\item $u_{jk} = u_{kj}^{-1}$.
\end{itemize}
Then we choose a reference site hamiltonian $h^0_j$ (whose groundstate is a product state $\otimes_j\ket{0_j}$) 
and take
$$ h_j = \prod_{l \in v(j) } u_l^{ s_l }  h^0_j  \prod_{l \in v(j) } u_l^{ -s_l }  + h.c.$$
and $P_j = {1\over n} \sum_{k=0}^{n-1} A_j^k$
is the projector onto $A_j= 1$.
Again the product over links $\prod_{l \in v(j)}$ is 
taken with the links pointing out of the site.

These terms commute\footnote{
Let $ -l$ denote the link $l$ traversed in the opposite direction,
so $ s_{-l} = (-s_l)_n, u_{-l} = u_l^{-1} $.
$(a)_n$ denotes $ a$ modulo $n$.
The version of 
\eqref{eq:hs-does-not-matter} where we 
keep track of $s$ mod $n$ (and regard the links as outgoing from the site $s$) is:
\bea \label{eq:hs-does-not-matter-N} 
{\cal B}_p h_s 
&=& \( \sigma_{l_1}^x \sigma_{-l_2}^x u_{l_1} u_{-l_2}...\) \( u_{l_1}^{s_{l_1}} u_{l_2}^{s_{l_2}} ... \) 
h_s \( u_{l_1}^{s_{l_1}} u_{l_2}^{s_{l_2}} ... \)^\dagger
\cr 
&=& u_{l_1} u_{-l_2} u_{l_1}^{(s_{l_1}-1) } u_{l_2}^{ ( s_{l_2} + 1 ) }   ..  h_s
.. u_{l_1}^{-s_{l_1}+1 } u_{l_2}^{-s_{l_2} - 1  }     \sigma^x_{l_1} \sigma^x_{-l_2} .. 
\cr &=&( u_{l_1} u_{l_2} )^{1-1} 
\underbrace{  u_{l_1}^{s_{l_1} } u_{l_2}^{  s_{l_2}   } ..  h_s 
.. u_{l_1}^{-s_{l_1} } u_{l_2}^{-s_{l_2}   }  }_{= h_s} 
\underbrace{ u_{l_1} u_{-l_2} ..      \sigma^x_{l_1} \sigma^x_{-l_2} .. }_{= {\cal B}_p}
\cr &=&  h_s {\cal B}_p .
\eea
} 
and the groundstate is a uniform sum of 
closed $ \IZ_n$ string nets occupied by the $\IZ_n$ state 
$$ \prod u_l^{s_l}  \otimes_j \ket{0_j} .$$
%[ should we write the actual full groundstate of the monster here, ]T\\
That is, the groundstate of the full monster is (up to normalization)
$$\ket{\gs}  =  \sum_C \prod\limits_{l \in C} \sigma^x_l u_l  
\Big(
\ket{0}\otimes \prod\limits_j \ket{0_j} \Big). $$
where, $C$ is closed $\IZ_n$ net and $\ket{0}$ is the state with no strings. 
Note that for $n>2$ we have {\it junctions}.

\section{Cluster states as 1d SPT states} 
\label{sec:1dreview}

Here we provide an example of a nontrivial 1d system
for which we know a circuit description meeting the demands above.
In the quantum information literature, these states 
are called {\it cluster states} or {\it graph states}.  
They optimize various measures of multipartite entanglement
and are the basis of the measurement-based quantum computing scheme.
For some further pedagogical discussion,
see chapter 10 of \cite{Rieffel-Polak}.

Consider an open chain of $N$ qbits with
\be\label{eq:cluster-h} h \equiv  -  \sum_{i=2}^{N-1} Z_{i-1} X_i Z_{i+1} . \ee
This hamiltonian is a sum of commuting terms and has a $\gG \equiv \IZ_2 \times \IZ_2$ symmetry 
generated by
$$ \mathsf{g}^{e/o} = \prod_{i,~\text{even/odd}} \XX_i .$$
The operators \cite{2013arXiv1307.4092B}
$$ \Sigma^x_L \equiv X_1 Z_2,~~
\Sigma^y_L \equiv Y_1 Z_2,~~
\Sigma^z_L \equiv Z_1 $$
satisfy the $\gSU(2)$ algebra, commute with the hamiltonian\footnote{For this property, it is important 
that we do not include the first term $ X_1 Z_2$, which is a
an external field for the effective edge spin.}
and do not commute with $\gG$.
The same statements apply to the other end:
$$ \Sigma^x_R \equiv Z_{f-1} X_f,~~
\Sigma^y_R \equiv Z_{f-1} Y_f,~~
\Sigma^z_R \equiv Z_f $$
All the states of the chain are therefore fourfold degenerate.
No perturbation which preserves $ \gG$ can split this degeneracy, so the symmetry protects the nontriviality of the state.  
%{\cobl (Notice that if the number of sites $f$ is odd, then only $g^o$ matters for protecting the edge states.  
%What's up with that?)}

A useful description of the ground states is obtained as follows.
First observe (\eg~\cite{2012Levin-Gu}) that 
$$ h  = - \UU \sum_i X_i \UU^\dagger $$
where 
%$$ \SS_s \equiv \prod_{i} e^{ { \pi \ii \over 4 } ( 1 - Z_i Z_{i+1} )  (-1)^i }. $$
\be\label{eq:cluster-unitary} \UU \equiv \prod_{i} {\bf CZ}_{i,i+1} 
\equiv \prod_i e^{ { \pi \ii  \( { 1 - Z_i  \over 2 } \) \( { 1 - Z_{i+1} \over 2 } \) }}  . \ee
%e^{ { \pi \ii \over 4 } ( 1 - Z_i Z_{i+1} )  (-1)^i }. $$
(For PBC, the product may run $i=1..N$, with $N+1 \equiv 1$; 
for the open chain it is $ i = 1..N-1$.)
There is some ambiguity in the form of $\UU$, in the form
of on-site unitary rotations of the link unitaries: 
$$ u_{i,i+1}  \to v_i^\nd w_{i+1}^\nd u_{i, i+1} w_{i+1}^\dagger v_i^\dagger  .$$
For example, the form of the link unitaries used in \cite{2012Levin-Gu} is 
$\tilde u_{i,i+1} = e^{ { \pi \ii \over 4 } ( 1 - Z_i Z_{i+1} )  (-1)^i }$.

\def\CZ{{\bf CZ }}
%\cor  It is useful to note that  $\SS_s = \prod_i {\bf CZ}_{i,i+1} $ up to phases.   \cob    

A groundstate is obtained by acting with $\UU$ on a groundstate of the symmetric trivial paramagnet
\be\label{eq:open-chain} h^0 = - \sum_{j=2}^{N-1} X_j  .\ee
For an open chain as in \eqref{eq:open-chain}, this hamiltonian is independent of the first and last spins,
so we obtain the four states
$$ \ket{\alpha_1, \alpha_N} = \UU \left(\ket{\alpha_1} \otimes_{i=2}^{N-1} \ket{\rightarrow}_i \otimes \ket{\alpha_N}\right) $$
where $ X_{1,N} \ket{\alpha_{1,N}}  = \alpha_{1,N} \ket{\alpha_{1,N}}$.
These states are eigenstates of 
$ \Sigma^x_{L, R}$ with eigenvalues $\alpha_{1,N}$.
\footnote{It is possible to show more directly that 
$$ \UU  \otimes_i \ket{\alpha_i}_i $$
is an eigenstate of $h$
using the fact that 
$$ X_i   \CZ_{i,i+1}
%e^{ {\ii \pi \over 4 }(1-  \ZZ_i \ZZ_{i-1} ) }
%=  e^{ {\ii \pi \over 2 } \ZZ_i \ZZ_{i-1} }  e^{ {\ii \pi \over 4 }(1-  \ZZ_i \ZZ_{i-1}  )} \XX_i 
=  Z_{i+1}    \CZ_{i,i+1}X_i , ~~~
 X_{i+1}   \CZ_{i,i+1}
%e^{ {\ii \pi \over 4 }(1-  \ZZ_i \ZZ_{i-1} ) }
%=  e^{ {\ii \pi \over 2 } \ZZ_i \ZZ_{i-1} }  e^{ {\ii \pi \over 4 }(1-  \ZZ_i \ZZ_{i-1}  )} \XX_i 
=  Z_{i}    \CZ_{i,i+1}X_{i+1} .  
 $$}

For closed chains, the Hamiltonian 
$$h = - \UU^\nd \sum_{j=1}^{N} X_j \UU 
= - \sum_{j=1}^N Z_{j-1} X_j Z_{j+1} 
$$ (with periodic boundary conditions, $N+1 \simeq 1$) has a unique ground state
$$ \UU  \otimes_j \ket{\rightarrow_j }  =\sum_{Z\text{-basis~states,} z = \pm }(-1)^\text{$\half$number of domain walls$(z)$} 
 \prod_{j=1}^N z_j \ket{z}. $$

% \subsection{Cluster states as SPTs for time reversal}
% 
% \label{sec:time-reversal-sym-of-cluster}

Another symmetry of the solvable model is time reversal symmetry.  That is,
$$ \CT = k \otimes \prod_j X_j $$
(where $k$ is complex conjugation)
is an antiunitary symmetry of the cluster hamiltonian \eqref{eq:cluster-h}.
%{\cor ...}
Notice that the individual link unitaries $u_{\vev{jk}} = \CZ_{jk}$ 
transform as 
$$ \CT: \CZ_{jk} \mapsto - Z_j \CZ_{jk} Z_k ~.$$
A chain of link unitaries only transforms at the endpoints
$$ \CT: \prod_{j =1}^N \CZ_{j,j+1} \mapsto  (-1)^{N-1} Z_1  \prod_{j =1}^N \CZ_{j,j+1} Z_N .$$
A closed circuit 
therefore maps to itself up to a sign.
%with an even number of sites is therefore invariant.

%\cite{PhysRevLett.100.167202}

\subsection{Stability of the edge states}
\label{sec:stability}

The degeneracy of the ground states is protected by the symmetry; no perturbation of the chain 
hamiltonian which preserves the $\gG$ can mix the states forming the doublet of the edge spin $\gSU(2)$ at one end. The edge states of an SPT form irreducible projective representations of the symmetry group. 
The irreducible property means that they do not mix with other states under application of any elements of the group,
and further, that only the elements of the group mix them with each other.
Therefore, an operator which mixes them cannot commute with the whole symmetry group.
So only non-symmetry-preserving perturbations can lift the degeneracy.

Let us illustrate this general statement explicitly in the example of the cluster model.
The cluster model supports two states at each edge which we can label as $\ket {\up},\ket\down$. 
For an open chain with an odd number of sites (a chain with an even number of sites can be analyzed similarly), the action of the symmetry generators on these states is given by
\begin{align}
 \mathsf g_o\ket{\up/\down} &= \pm \ket{\up/\down} & \mathsf g_e\ket{\up/\down} &= \ket{\down/\up} .
\end{align}
A symmetry preserving perturbation $\hat O$ satisfies $\mathsf g_e^\dagger \hat O \mathsf g^\nd_e = \mathsf g_o^\dagger \hat O \mathsf g^\nd_o = \hat O$. Then it follows that 
\be
\begin{split}
 \bra\up \hat O \ket\up &= \bra \up \mathsf g_e^\dagger \hat O \mathsf g^\nd_e \ket\up = \bra\down \hat O \ket\down \\
 \bra\up \hat O \ket\down &= \bra\up \mathsf g_o^\dagger \hat O \mathsf g^\nd_o \ket\down = -\bra\up \hat O \ket\down  = 0
\end{split}
\ee

As a result, perturbing by the operator $\hat O$ will not split the degeneracy between the edge states. The operator which mixes the states $\ket\up,\ket\down$ is the symmetry generator $\mathsf g_e$, a nonlocal operator spanning the whole chain. Therefore the splitting of the states by a local perturbation is exponentially small in the system size.

\section{An exactly solvable spin liquid 
with spinons}

\label{sec:cluster-snake}
Exactly solvable models such as Kitaev's toric code and honeycomb models 
\cite{2006AnPhy.321....2K,2003AnPhy.303....2K}
have played an important role in our understanding of spin liquids\footnote{A useful review of models of spin liquids for our purposes is \cite{2013NJPh...15b5002G}.
}.
However, these models have no essential symmetries,
in the sense that 
although the solvable limit of the models do have various global symmetries,
those symmetries do not act nontrivially on the quasiparticles.

On the other hand, electrons carry spin and their interactions often preserve spin rotation invariance.
Real spin liquids, when we find them, may have excitations which carry spin quantum numbers,
namely {\it spinons}, and it would be useful to
have solvable examples of this phenomenon.

To our knowledge there is so far no known exactly solvable model
of a spin liquid with spinons, in this sense, above one dimension.
In 1d, the spin-half-odd Heisenberg chain provides an example.

There are well-known models of spin liquids, which have also played an important role in the history of the subject, whose exact {\it groundstate} is known \cite{Rokhsar:1988zz, 2001PhRvL..86.1881M}.
That is, the associated parent hamiltonians are {\it frustration free}:
each term independently annihilates the groundstate.
Some frustration-free models exhibit phenomena similar to what we describe below
\cite{2010arXiv1012.4470Y,2014PhRvB..90d5142H,2014PhRvB..89q4411L}.
In particular,
%\cor (Say more about the symmetries of the quantum dimer model.  
\cite{2005PhRvB..72f4413R} discusses a quantum dimer model which preserves spin rotation symmetry.

The interplay between global symmetry and topological order lies at the heart of the study of symmetry enriched topological (SET) phases. 
In a model with both topological order and global symmetries, anyons carry fractional quantum numbers. The type of fractionalization characterizes a particular SET phase \cite{2013PhRvB..87j4406E, 2013PhRvB..87o5115M}.

In this section, we show that the model 
defined by the hamiltonian 
\eqref{eq:general-snake-hamiltonian}
with the cluster state link unitaries \eqref{eq:cluster-unitary}
represents a gapped spin liquid with spinons.
In the example we will study, the elementary degrees of freedom
are effectively integer-spin excitations, 
while the quasiparticles in the spin liquid phase have half-integer spin.

\subsection{Unitary symmetry}
\def \CZ{{\bf CZ}}
Explicitly, the Hamiltonian is given by 
\be
\label{eq:cluster-snake-hamiltonian}
\HH = -\sum_i A_i - \sum_p \CB_p - \sum_i X_i\prod_{\vev{i|j}}Z_j^{s_{ij}}\left(\frac{1+A_i}{2}\right)
\ee
The cluster hamiltonian \ref{eq:cluster-h} for a {\it closed} chain has a $\gG = \IZ_2 \times \IZ_2$ symmetry 
generated by $\mathsf g^{e/o}$.
This is respected by the star term, which doesn't involve the site variables.
The plaquette term in $\HH$ does not obviously respect this symmetry, 
since it involves products of $Z$s on the two sublattices.
A crucial fact here is that any bipartite lattice has an even number of sites $2n$
around every plaquette.
\bea (\mathsf g^o)^\dagger S_\Box  \mathsf g^o 
&=& X_1 X_3 \cdots X_n
\prod_{j =1}^{2n} 
%e^{{\ii \pi \over 4} ( 1 - Z_j Z_{j+1} ) (-1)^j }
\CZ_{j, j+1}
X_1 X_3 \cdots X_n
%\cr 
%&=& \prod_{j } \CZ_{j, j+1}
%e^{{\ii \pi \over 4} ( 1 + Z_j Z_{j+1} ) (-1)^j }
%\cr &=& %\prod_{j} e^{{\ii \pi \over 2} Z_j Z_{j+1}  (-1)^j }
%S_\Box 
\cr 
&=& (Z_2 Z_4  )(  Z_4 Z_6)\cdots ( Z_{2n} Z_2   ) S_\Box = S_\Box
\eea
so it is actually invariant.
 
\cob

The site hamiltonian $h_i =X_i\prod_{\vev{v|j}}Z_j^{s_{ij}}$ on an {\it open} string is not $\IZ_2 \times \IZ_2$ symmetric;
it does not commute with the endpoint terms.
$$ [ \mathsf g^e, X_1 Z_2 ] \neq 0 , ~~[\mathsf g^\alpha , Z_{f-1} X_f] \neq 0  $$
Here $ \alpha =$ odd/even for $f$ even/odd respectively.
%This means that when acting on configurations of the 
%link variables corresponding to open strings, 
Multiplying the site hamiltonian by a projector onto locally closed strings guarantees that only symmetric terms appear.
\def\gh{{\mathfrak{h}}}
\def\gz{{\mathfrak{z}}}
\def\gx{{\mathfrak{x}}}
\def\sameends{\circ---\circ}
\def\diffends{\circ---\bullet}

\subsection{String operators and anyons}
\label{sec:String-operators-and-anyons}

The magnetic string operator is unmodified relative to the toric code: $M_{\check C} = \prod_{l \perp \check C} \sigma_l^z $.
Notice that if the curve in the dual lattice $\check C$ just goes around one site
in the primal lattice, this reproduces the star operator, as usual.
For closed $\check C$, this operator commutes with $ \HH$;
if $ \check C$ ends, $M_{\check C}$ violates the plaquette operators at the endpoints.
This will mean that the anyons have the same statistics as in the toric code since 
only the link variables can participate in the commutator.

The demand that $W_{C = \partial p}  = {\cal B}_{p }$ 
suggests that that the electric string operator is 
$$ W_C = \prod_{l \in C} \sigma_l^x \CZ_l . $$
This operator indeed commutes with $ \HH$ for closed curves.

Consider next the operator $W^{1,f}_C$ associated with an open string $C$ with endpoints $1,f$:
$$ W_C \equiv \prod_{l \in C} \sigma^x_l  \CZ_{12} \CZ_{23} ... 
\CZ_{f-1,f}. $$
Acting on the groundstate of the snake monster, this violates the star constraint at sites $1,f$
(that is, $  W_C A_{1,f} = - A_{1,f}  W_C$).

We may modify our string operator by decorating it with site operators localized at the endpoints.
Thus, we have found {\it four} states associated with each open string, 
a two-dimensional Hilbert space for each endpoint spanned by $\{W^{a,b}_C\ket{\textrm{gs}}, Z_a W^{a,b}_C \ket{\textrm{gs}}\}$. These states are eigenvectors of the end-point-site hamiltonians $h_{a,b}$ with eigenvalue $+1$ and $-1$ respectively, and we will occasionally refer to them as $\ket{+}$ and $\ket{-}$. The projector onto $A_i =1$ annihilates these states, so they are degenerate.

To see that these states form an orbit of the $\IZ_2\times\IZ_2$ symmetry we act on them with the generators. Straightforward calculations involving the algebra of $X$ and $\textbf{CZ}$ yield

\be\begin{split}
\mathsf g_o W_{e,e}\ket{\textrm{gs}} &= Z_1 Z_f W_{e,e} \ket{\textrm{gs}} \\
\mathsf g_e W_{e,e}\ket{\textrm{gs}} & = W_{e,e}\ket{\textrm{gs}}
\end{split}
\ee
where the subscripts on the open string operator $W$ indicate on which sublattice the string begins and ends.
Furthermore, $\mathsf g_e$ anticommutes with $Z_i$ when site $i$ is on the even sublattice. We can therefore summarize the action of the symmetry on the anyon states as follows:

\begin{center}
\begin{tabular}{|c|c|c|}
\hline 
 $\bullet = e,\circ = o$& $\mathsf g_e$ & $\mathsf g_o$ \\ \hline \hline
 $\bullet$ - - - $\bullet$ & $\bf{z\otimes z}$ & $\bf{x\otimes x}$ \\ \hline
 $\bullet$ - - - $\circ$ & $\bf{z\otimes x}$ & $\bf{x\otimes z}$ \\ \hline
 $\circ$ - - - $\bullet$ & $\bf{x\otimes z}$ & $\bf{z\otimes x}$\\ \hline
 $\circ$ - - - $\circ$ & $\bf{x\otimes x}$ & $\bf{z\otimes z}$ 
 \\
 \hline
\end{tabular} 
%\label{table:endpoint-action}

{\footnotesize Table: Action of $g^{o/e}$ on the anyons.}
\end{center}
This table encodes the projective representation of $\IZ_2\times \IZ_2$ 
furnished by each anyon. In a system with periodic boundary conditions, a single anyon is not a physical state; they always come in pairs. We see that while an endpoint individually represents the algebra $\mathsf g_e \mathsf g_o \sim -\mathsf g_o \mathsf g_e$, the symmetry generators act linearly on the entire many-body wavefunction as they must. 
 
%\subsubsection{Characterizing the SET order of the snake monster}

In summary, the structure of topological order (the quasiparticle labels and statistics) in this model is the same as the toric code, but the e particles form doublets. 
The argument for the robustness of the degeneracy given for the 1d model 
carries over directly to the snake monster.
In \S\ref{sec:snake-monster-stability} 
we study symmetric perturbations of this hamiltonian 
and show explicitly that the characteristic feature
(the projective representation of $\mathsf g_e$ and $\mathsf g_o$ on the 
quasiparticles) is preserved.

It would be interesting to apply the methods of 
\cite{2014arXiv1410.4540B}
to more precisely characterize the SET order of the cluster snake monster in the case of $d=2$.
%Twisted boundary conditions on the strings.
It would also be interesting to gauge the $ \IZ_2\times \IZ_2$ symmetry; 
this can be done maintaining solvability and 
produces a model with topological order (and no global symmetry) whose spectrum
is characteristic of the original SET.

\subsection{Time-reversal-invariant cluster snake}

\label{sec:snake-SET}

The same model
also produces a solvable representative
of an SET protected by time-reversal symmetry.
Of course, the general $\CT$-preserving perturbation
will be different than that preserving the unitary symmetry described above.

In this case, the string endpoints are Kramers' doublets,
a projective representation of the antiunitary symmetry $\CT$,
in the sense that $\CT^2=-1$ on the states of the endpoints.
We can see this as follows, using $ \CT: \CZ_{12} \mapsto - Z_1 \CZ_{12} Z_2$.
Consider a open string excitation with an endpoint labelled by 1, 
\be\label{eq:endpoint-cluster}
W_0 \ket{\gs} = \prod\limits_{j=1}\sigma_{j,j+1}^x 
%{\mathfrak{X}_{j,j+1}}   {\mathfrak{X}_{j+1,j+1}} 
\CZ_{j, j+1} 
\ket{\gs} .
\ee
This 
is an excited state which 
violates the star constraint at the end site, 
and is annihilated by the site term at the end site (whereas sites participating 
in closed strings are eigenstates with eigenvalue $-1$).
A degenerate and orthogonal state which is the Kramers' doublet partner 
is obtained by acting upon \eqref{eq:endpoint-cluster} with time-reversal:
$$
W_1\ket{\gs} =   \CT W_0 \ket{\gs} = (-1)^{f-1} Z_1 Z_f W_0 \ket{\gs}.
$$
The states are orthogonal because 
$$ 
\bra{\gs } W_1^\dagger W_0 \ket{\gs}  = (-1)^{f-1}  \bra{\gs} Z_1 Z_f \ket{\gs}
$$
which vanishes by symmetry\footnote{
To see this more explicitly, recall that the groundstate is a 
uniform superposition of closed string configurations.
$\CZ$ and $Z$ commute, so the expectation value of $Z$ in an arbitrary closed string configuration is
$$ \vev{ \rightarrow_0 \rightarrow_1 \rightarrow_2 | \CZ_{01}^{-s_{01}} \CZ_{12}^{-s_{12}}   Z_1 
\CZ _{01} ^{s_{01}} \CZ_{12} ^{s_{12}}   | \rightarrow_0 \rightarrow_1 \rightarrow_2  } 
= \vev{ \rightarrow  | Z_1 |\rightarrow } = 0 . $$
}. 

Notice that a second action of $\CT$ on 
the endpoint $ Z_1 W_0 \ket{\gs} $ 
flips the sign of $ Z_1$, so indeed $ \CT^2 =-1$.

In \S\ref{app:cluster-Zn} we generalize the construction to $\IZ_N$ cluster states and  
describe the snake monster on a $\IZ_N$ string net. 
%$$ \CT: \prod_{j =1}^f \CZ_{j,j+1} \mapsto  Z_1  \prod_{j =1}^f \CZ_{j,j+1} Z_f .$$

%AKLT was here

\section{Exactly solvable models of topological paramagnets}

\label{sec:3d-paramagnets}

Not all states which are equal-magnitude superpositions of closed string configurations
have topological order.  
Finding a hamiltonian with such string-net groundstates 
which does {\it not} have topological order
requires a local condition which prevents the strings from winding around noncontractible cycles.
One way to do this is if the curves are all boundaries of a region.
This is what happens in the quantum Ising paramagnet, 
where the groundstate of $ h = - \sum_j X_j $ can be written in the $z$-basis
as a uniform superposition of closed loops
which are the boundaries of domain walls
 (A similar statement can be made about the 
groundstate of the Levin-Gu model \cite{2012Levin-Gu}, 
which involves an additional phase which counts the number of walls.)

For the case of 3d lattices, 
\cite{wns2015}
provides a beautiful mechanism to accomplish this goal.
%construction of a model which is actually trivial.
As we review next (and elaborate in \S\ref{app:pure-loop}), it provides a mechanism for destroying the topological order of a 
model with string condensation.
In the appendix \S\ref{app:ZN-pure-loop} we also generalize the construction from $\IZ_2$ to $\IZ_N$.

\subsection{Pure loop model review}

%\S\ref{sec:Z4snakes}.

Following \cite{wns2015}, 
consider two interpenetrating cubic lattices $A$ and its dual lattice $B$. 
This means that the vertices of $A$ are in the centers of the cubes of $B$ (and vice versa)
and each link of $A$ pierces a plaquette of $B$ (and vice versa).
Put qbits on the links of both, and denote the associated Pauli operators
for links of $A,B$
by $ \sigma, \tau$ respectively.
Consider 
$$ \HH_\text{linking} \equiv 
- J \( \sum_{p \in A} \CF_{Ap} + \sum_{p \in B} \CF_{Bp} \)  $$
where 
$$ 
 \CF_{Ap} \equiv \tau_p^z \prod_{ l \in p } \sigma^x_l , ~~~~~~~~~~~~
  \CF_{Bp} \equiv \sigma_p^z \prod_{ l \in p } \tau^x_l , ~~
$$ 

Claims from \cite{wns2015}:
\begin{enumerate}
\item All the $\CF$s commute with each other.
\item The condition $\CF_{Ap} = 1$ says
that if there is electric flux on the B-link $p$ ( $\tau_p^z=-1$)
then there is magnetic flux on the A-plaquette $p$ ($  \prod_{ l \in p } \sigma^x_l $).  
So this hamiltonian glues the electric flux lines of the $A$ gauge field
to the magnetic flux lines of the $B$ gauge field.  
%Very nice and useful.
This
is a lattice realization of the $B\wedge F$ term.

\item We don't need to add star operators to $\HH_\text{linking}$ because
they are products of the $\CF$s.  That is, if we have $\CF=1$ for all $p\in A,B$
then automatically $ \prod_{l \in v} \sigma^z_l = 1 $ for all vertices $v \in A$
and $ \prod_{l \in v} \tau^z_l = 1 $ for all vertices $v \in B$.
More explicitly, the star operator for a site $v \in A$ (which is at the center of 
a cube $v \in B$) is
$$ \prod_{l \in v} \sigma^z_l = \prod_{p \in \partial v } \CF^B_p $$
where by $\partial v$ we mean the six faces which bound the cube $v$.

\item The unique groundstate on any manifold is
\bea 
 \ket{\gs} &=& \sum_{\begin{matrix}\tiny{C_A, C_B}\\ \tiny \text{contractible mod 2}\end{matrix}} (-1)^{\ell(C_A, C_B) } \ket{C_A} \otimes \ket{C_B}  \cr 
 \label{eq:membranes}
 &=& \sum_{C_A} \sum_{\CM \subset \Omega_2(A)| \partial \CM = C_A }
 \ket{C_A} \ket{\CM}~.
\eea
where $\ell(C_A, C_B)$ is the linking number of the two sets of loops.
In the last expression $ \Omega_2(A) $ denotes plaquettes of $A$, 
and 
$$ \ket{\CM} \equiv \left |  \tau^x_p = \begin{cases} -1, ~\text{if} ~ p \in \CM, \cr 
+1, ~\text{else} ~ 
\end{cases} \right \rangle$$
The point of this last expression is that it makes clear
that the loops $C_A$ must be homologically trivial (mod two) 
since they are boundaries of the membranes $\CM$ specifying the state of the B-lattice variables.

\item If the lattice has a boundary, this model has surface topological order, 
the same as the toric code.  
The $e$ particles are the ends of the electric strings on the $A$ lattice,
and the $m$ particles are the ends of the electric strings on the $B$ lattice.
These particles are deconfined bosons which are mutual semions.  
String operators which create them in pairs 
can be written as shown in fig 5 of \cite{wns2015}.
More explicitly,
for smooth boundary conditions on both $A$ and $B$ lattices, the $e$ particle is created 
on the A sublattice by 
$\prod_{l \in L} \sigma^x_l $
where $L$ lies on the boundary of A. 
$\prod_{l \perp \check L}\sigma^z_l  \tau^x_{p_l}$
creates the magnetic excitations on the $A$ 
sublattice boundary.
We see that this involves an electric string on the 
boundary of the $B$ sublattice.

\item This model does not have topological order and is actually
adiabatically connected to a product state as shown explicitly
in \cite{wns2015}.

\end{enumerate}

\subsection{Cluster snake paramagnet}

In order to build solvable three-dimensional SPTs
starting from the pure loop construction, 
we will decorate the strings with extra degrees of freedom living on the sites.
As we will explain, 
it is necessary to decorate both sublattices; adding qbits 
on one of the sublattices alone is not enough to generate a distinct phase. 
One way to see this is to use the membrane representation for the sublattice without the decorations. 
Then it is possible to adiabatically contract the membranes without breaking any symmetry of the site variables.

So, now add qbits on the sites of both lattices.  
%(We will need to discuss what happens if we put them on only one sublattice.  The model should be trivial
%since then you can use the membrane representation for the sublattice without the decorations,
%and adiabatically contract the membranes without breaking any symmetry of the site variables.)
We replace the magnetic flux operators 
$B_{Ap} \equiv  \prod_{ l \in p } \sigma^x_l $
with snake monster operators: 
$$ {\cal B}  _{Ap} \equiv \prod_{ l \in p } \sigma^x_l  U_{\partial p} $$
where $ U_{\partial p}$ is defined as in the previous sections on snake monsters.
That is, consider instead: 
$$ \HH_\text{snake-linking} \equiv 
- J \( \sum_{p \in A} \CF_{Ap} U_{\partial p} + \sum_{p \in B} \CF_{Bp} U_{\partial p} \)  
- \sum_v h_v 
$$
where 
$$ h_v \equiv  X_v \prod_{\vev{ v|w} } Z_w^{s_{vw} }\left(\frac{1+A_v}{2}\right) $$
where $s_{vw} \equiv \half ( 1 - \sigma^z_{vw} ) $ if $vw$ is an A-lattice link or
$s_{vw} \equiv \half ( 1 - \tau^z_{vw} ) $ if $vw$ is a B-lattice link.

These terms still commute.  
The star operators for the 
A and B sublattices are still products of the flip operators.
This model still does not have topological order,
since in the groundstate, the electric strings on each sublattice 
are boundaries of membranes in the dual lattice.
The unique groundstate is 
$$ \ket{\gs} = \sum_{\begin{matrix}\tiny{C_A, C_B}\\ \tiny \text{contractible mod 2}\end{matrix}} (-1)^{\ell(C_A, C_B) } \ket{C_A} \otimes \ket{C_B} \otimes  U_{C_A \cup C_B} \ket{\rightarrow}^{\otimes v} . $$

With smooth boundary conditions on both A and B lattices, 
%we find a copy of the 
the surface is gapped and symmetric.
The surface topological order is the same as for the pure loop model with the crucial difference that the anyons form doublets. Due to the binding of electric and magnetic flux, both $e$ and $m$ now form projective representations of the global symmetry.

Other choices of boundary conditions are possible.
However, the all-smooth boundary conditions are most convenient 
\cite{WangPotterSenthil, PhysRevB.88.115137, Metlitski:2013uqa, Bonderson:2013pla,
2013PhRvX...3d1016F, 2014arXiv1403.6491C, 2013arXiv1306.3286M}
because they produce a nondegenerate, gapped, symmetric groundstate 
(when the boundary is simply connected).
%[cite papers about anomalous TO.]}
%all the anyons 
%are doublets which are projective representations of $ \IZ_2 \times \IZ_2$.

\subsection{Nontriviality of the snake paramagnet}

If we put cluster snakes on {\it only} the A sublattice 
the model is trivial:
it can be shown to be adiabatically connected to a product
state while preserving all symmetries
by the string tension deformation described in \cite{wns2015}: 
Deform the flipper 
$$\CF^A_p \to \CF^A_p(\gamma)\equiv
\cosh^{-1}\gamma\( \tau_p^z \prod_{l \in p} \sigma^x_l  + \tau_p^x  \sinh \gamma \) ~;$$
 these still commute, but interpolate to a product groundstate as $ \gamma \to \infty$.
If B were decorated as well, then this operator would break the B sublattice symmetry.

{\bf Cluster paramagnet with $\CT$ symmetry.}
% {\cobl Here is what's nontrivial about the surface topological order 
% that results when we decorate both A and B lattice electric strings 
% of the pure loop construction
% with {\it time-reversal} SPTs.}

Decorating both A and B lattice electric strings of the pure loop construction with time-reversal SPTs results in an `anomalous' surface topological order, a characteristic feature of three dimensional bosonic SPTs.
The spectrum of quasiparticles is:

\begin{table}[h!]
%\caption{default}
\begin{center}
\begin{tabular}{|c||c|c|}
\hline 
quasiparticle   & self-statistics &  time-reversal property \\
\hline \hline
e &    B  & 1/2  \\
m  &  B  &  1/2  \\
$\eps$ &  F  &   0 \\
\hline
\end{tabular}
\end{center}
\label{default}
\end{table}%

In this table, a $1/2$ denotes a Kramers' doublet.
The crucial property here 
is that the fermionic quasiparticle is the only time-reversal singlet.
This spectrum has the consequence that 
the surface topological order cannot be destroyed while preserving time-reversal symmetry.

To see this, recall that 
destroying topological order in 2d $\IZ_2$ gauge theory
requires condensing some anyon.  
Condensing $e$ is higgsing and condensing $m$ is confinement;
the resulting two phases are adiabatically connected \cite{FS7982}.
In the model we've constructed, 
$e$ cannot condense in a $\CT$-symmetric way because it is a Kramers' doublet.
That is, condensing any one $e$ particle will break the time-reversal symmetry.
This much can be realized (and we did realize it in \S\ref{sec:snake-SET} above)
intrinsically in two dimensions.
However, when we decorate the electric strings on {\it both} A and B lattices,
the ends of the B-lattice electric strings
behave as the $m$ particles, 
which are therefore also Kramers' doublets.
They can therefore also not condense in a $\CT$-symmetric way.
Finally, as usual, $\epsilon = em$ can't condense because it's a fermion.  
Condensing pairs of these objects doesn't destroy the topological order.

This proves that the edge of our 3d model
has no trivial gapped and symmetric edge,
and therefore represents a nontrivial SPT protected by $\CT$.
Since the surface quasiparticles
are Kramers' doublet bosons,
this is the state labelled $eTmT$ in the 
classification reviewed in
\cite{Senthil:2014ooa}.
The surface theory with this spectrum
is not `edgeable'.

To summarize: the snake monster produces a model where the electric defects are doublets. 
In the topological paramagnet, the magnetic defects become doublets by binding to electric defect doublets on the other sublattice. If we don't decorate sublattice A then the $m$ particle at a surface of sublattice B is a singlet under time reversal and can be symmetrically condensed, destroying the topological order. This further supports the previous assertion that it is necessary to decorate both sublattices to generate a distinct phase.

{\bf Cluster paramagnet with unitary symmetry.}
The cluster hamiltonian $ \sum_v h_v$ is also a $\IZ_2 \times \IZ_2$ SPT
state on a bipartite lattice.  
This requires both $A$ and $B$ to be bipartite.
If we put cluster snakes on both A and B sublattices (which are each bipartite)
the solvable model actually has a $ \IZ_2^4$ symmetry.  
The spectrum of excitations is much as in the table above, 
if we interpret the $ \half$ to mean a projective doublet of the $\IZ_2\times \IZ_2$
coming from the simultaneous spin flips on even and odd sublattices of both A and B.

Though it seems impossible to destroy this topological order symmetrically,
in the absence of time-reversal symmetry, the statistics of the anyons
can be changed by perturbations.  We have not settled the question of 
whether this phase is nontrivial as an SPT for unitary symmetry.
We note that the group cohomology classification
of \cite{PhysRevB.87.155114} 
has $ H^4(\IZ_2 \times \IZ_2, U(1) ) = \IZ_2 \times \IZ_2$.

\subsection{Relation to coupled layer construction}

It is instructive to ask about the relationship
between the above solvable model for the $eTmT$ state
and the coupled layer construction implied in \cite{WangPotterSenthil, PhysRevB.88.115137, Senthil:2014ooa}.  
A nice direct connection can be understood as follows;
it uses the 2d cluster snake monster of \S\ref{sec:snake-SET} in a satisfying way.

\begin{wrapfigure}[11]{r}{0.45\textwidth}
  \vspace{-55pt}
  \begin{center}
    \includegraphics[width=0.45\textwidth]{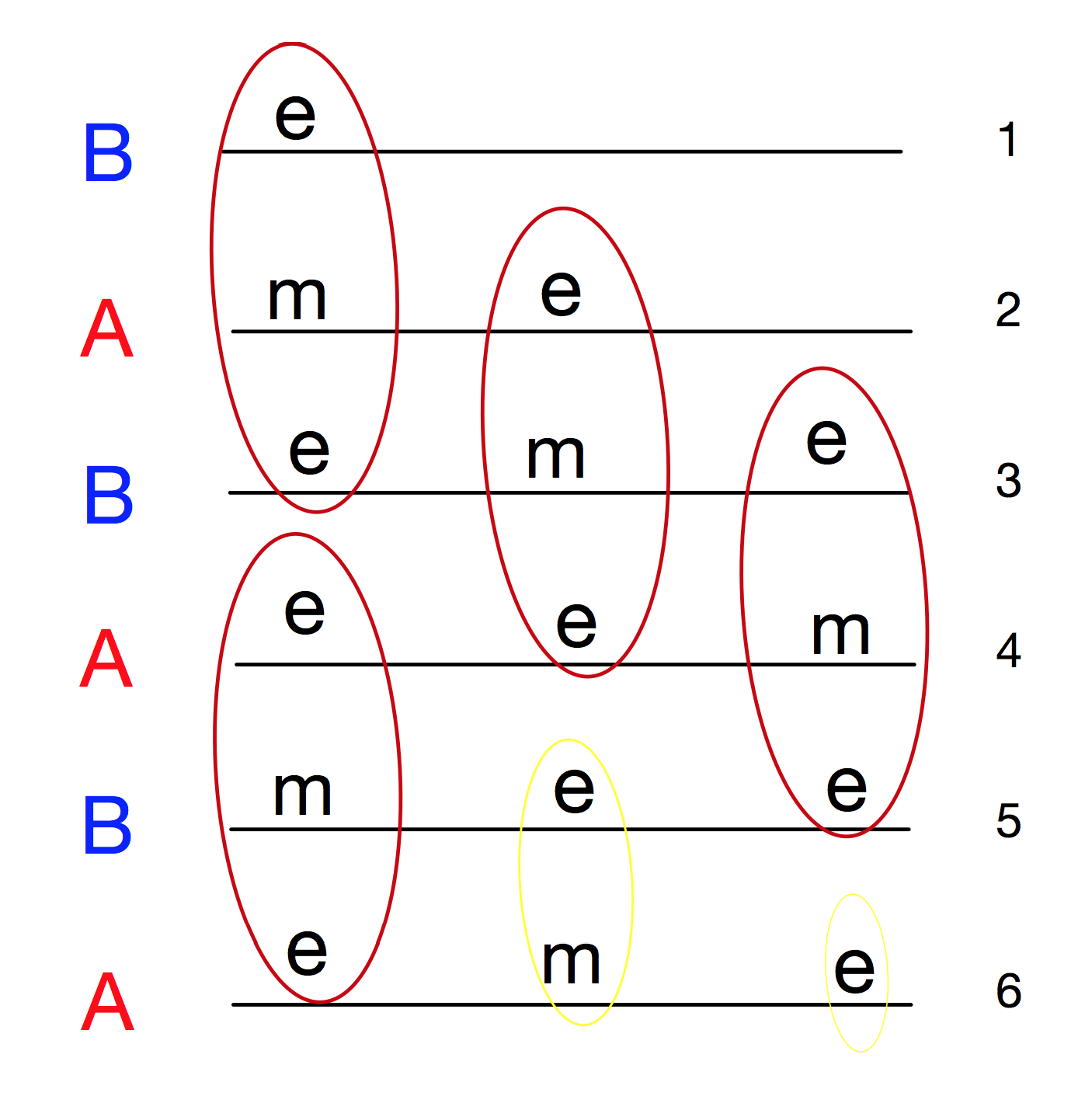}
  \end{center}
  \vspace{-20pt}
%  \caption{Sod.\label{fig:sod}}
%  \vspace{-20pt}
\end{wrapfigure}Consider first a collection of layers of ordinary 2d toric code as in the figure at right.
Condense the bosons $ b_n \equiv e_n m_{n+1} e_{n+2}$
(circled in red in the figure).
This higgses the gauge group of the layers of the same parity ($n$ and $n+2$)
to the diagonal $\IZ_2$ subgroup,
while at the same time confining the gauge group of the other-parity layer ($n+1$).
The electric flux lines of the even layers are attached to the magnetic flux lines 
on the odd layers and vice versa.
The bulk is trivial, since every layer is confined by one of the $b_n$ condensates.
This is therefore (a coarse-grained version of) the pure-loop construction; 
the A sublattice arises from the even layers and the B sublattice
from the odd layers.
At the surface is a copy of the ordinary toric code 
where 
the excitations at the surface 
which are mutually local with all the condensates
are the following (indicated in yellow in the figure).
The effective $e$ particle is the boson $e_\text{last}$ 
and the effective $m$ particle 
with which it is a mutual semion is the boson 
$m_\text{last} e_{\text{last}-1}$.
This is perfectly ordinary and trivial, as expected.

Now take instead layers of the 2d cluster snake monster of \S\ref{sec:snake-SET}:
the 2d toric code where the electric flux lines are decorated by a SPT of time-reversal.
This means that the $e$ particles in any layer are Kramers' doublets, $e_{\alpha = \up/\down}$.
Now we condense the Kramers' singlet bosons
$ b_n \equiv e_n^\alpha m_{n+1} e_{n+2}^\alpha$.
The bulk is again trivial.
The remaining surface excitations 
are now the Kramers' doublets 
$e_\text{eff}^\alpha = e_\text{last}^\alpha$ 
and 
$m_\text{eff}^\alpha = m_\text{last} e_{\text{last}-1}^\alpha$.

\section{Discussion}
\label{sec:discussion}
 
 The problem of finding circuit constructions of 1d SPTs 
meeting the demands listed in \ref{sec:snake-def} is an interesting one.
If such a circuit could be found for a single copy (or an odd number of copies)
of the Kitaev 
chain \cite{Kitaev:2000}, we would have a solvable gapped model in arbitrary dimension
with deconfined {\it non-abelian} anyons, 
along similar lines to the suggestion of \cite{Teo:2009qv}.

Although this is not a flat contradiction with the classification of particle statistics (since 
the information about the strings which end on these particles enhances the topology of the 
configuration space beyond that of particles \cite{Freedman:2010ak})
many attempts at such a construction 
\cite{Teo:2009qv, McGreevy:2011if, 2011PhRvB..84x5119F}
have failed to produce deconfined, gapped non-abelian particles in $d>2$, for interesting reasons.
In particular, strong evidence against this possibility 
from a low-energy field theory viewpoint
is given in \cite{McGreevy:2011if}.

Here the obstruction is the fact that a single copy of the Kitaev chain $h_1$
is a distinct phase from the trivial chain $h_0$
even in the absence of symmetry.
On the other hand, if one found the desired link unitaries which
relate the two, one could
then isospectrally interpolate between the two by $h_s = U^s h_0 U^{-s} , s \in [0,1]$.
In other cases, this is prevented by the fact that $U^s$ is not symmetric.

It would be interesting to find a sharp characterization of which 1d SPTs 
have such a description.

Here we have attached interesting
1d phases to
1d electric flux lines of 1-form discrete lattice gauge theory.
In a future publication we will show how 
to attach in a similarly explicit manner 
$p$-dimensional topological phases to the $p$-dimensional electric flux sheets
of $p$-form gauge theory.

\breakS

\vskip.2in
{\bf Acknowledgements}

We thank Dan Arovas and Shauna Kravec for many useful discussions.
JM would like to thank 
Isaac Kim and Brian Swingle for discussions which inspired the construction described here, 
and 
Maissam Barkeshli and
Ashvin Vishwanath for some very helpful comments.
This work was supported in part by
funds provided by the U.S. Department of Energy
(D.O.E.) under cooperative research agreement DE-FG0205ER41360.

\appendix

%\numberwithin{equation}{section}

\renewcommand{\theequation}{\Alph{section}.\arabic{equation}}

\section{Cluster states for $\IZ_N$}
\label{app:cluster-Zn}

Consider now an $N$-dimensional Hilbert spaces at the sites.
We will use the conventions
$$ X Z = \omega Z X , ~~X = \sum_n \ket{n} \bra{n+1} , ~~Z = \sum_n \omega^n \ket{n}\bra{n} . $$

A $\IZ_N$ generalization of control-$Z$ is 
$$ \CZ_{12} = \sum_{mn} \ket{mn} \bra{mn} \omega^{mn} ~,$$
which satisfies
\be\label{eq:CZ-X-algebra}
 \CZ_{12}^k X_2 = X_2 \CZ_{12}^k Z_1^{-k},\; \CZ^N=1 .
\ee

Consider also a $\IZ_N$ (bipartite) string net: 
that is, assign to the edges $\vev{ij}$ of a (bipartite) graph 
a configuration
of integers $s_{ij} $ mod $N$, satisfying
$$ s_{ij} = - s_{ji} $$
and
\be\label{eq:closed-N} \sum_{ \vev{ i|j } } s_{ij} = 0 , ~~\forall i \ee
-- the net flux into each site is zero, so the strings are closed.
(The notation $ \sum_{\vev{i|j}}$ means sum over neighbors $j$ of a fixed site $i$.) On a bipartite lattice, a canonical orientation for the links is pointing from the A sublattice to the B sublattice.

For each site, let $$ H_j = u^\dagger_j h^0_j u^\nd_j = -u^\dagger_j (X^\nd_j + X^\dagger_j)u^\nd_j = -X_j\left(\prod_{\vev{j|k}}Z_k^{s_{jk}}\right)^{\sgn(j)}+\textrm{h.c}$$
where $u_j = \prod_{\vec l \in v(j)}\CZ_{l}^{s_{l}}$ is a product of unitary operators on the oriented links in the vicinity of $j$ and $\sgn(j)$ is $+1(-1)$ for $j$ on the A(B) sublattice. These terms commute in the same way as the terms in the $\IZ_2$ cluster Hamiltonian.

The plaquette term in a $\IZ_N$ toric code is a product of alternating operators $\sigma^x$ and $(\sigma^x)^\dagger$ on the links bounding the plaquette. For example, on a square lattice we have $B_p = \sigma^{x }_N (\sigma^x_E)^\dagger \sigma^{x}_S (\sigma^x_W)^\dagger$. Under snake monsterification, this becomes $\CB_p = B_p \CZ^\nd_N \CZ^\dagger_E \CZ^\nd_S \CZ^\dagger_W$. The full Hamiltonian for the $\IZ_N$ theory  is then
$$ \HH = \sum_j 
\CV\(A_j\) + \sum_p \CV\(\CB_p \) + \sum_j H_jP_j $$
where $\CV(z)$ is a real-valued function of a phase $|z|^2 =1$
which is minimized when $z=1$.

The groundstate of this model is
$$ \ket{\gs} = \CN^{-1/2} \sum_C  (W^\nd_C+W^\dagger_C )   
\ket{0} \otimes \prod_j \ket{0_j} .$$
where $C$ is a closed $\IZ_N$ string-net and 
$W_C = \prod_{l \in C}  (\sigma^x_l \CZ_l)^{n_l} (\sigma^x_{l+1}\CZ_{l+1})^{-n_{l+1}}\ldots $ is the $\IZ_N$ dimensional analog of the string creation operator. The exponent $n_l$ is the multiplicity of link $l$ in the string-net $C$,
$\ket{0}$ is the empty link configuration, 
and $ \ket{0_j}$ is the groundstate of $ h^0_j$.

When the graph is bipartite, this model has a $\IZ_N \times \IZ_N$ symmetry
which is generated by 
$$ \mathsf{g}^{o/e} \equiv \prod_{j\in o/e} X_j .$$

A quasiparticle is obtained by acting on $\ket{\gs}$ by $W_L$ where $L$ is an open curve. 
Thus using \eqref{eq:CZ-X-algebra} we see that the successive action of 
$\mathsf{g}^{o/e}$  generates a set of degenerate, orthogonal quasiparticle states. For an endpoint labelled by $1$, 
there is a $N$-dimensional Hilbert space spanned by $\{Z_1^k W_L \ket{\gs}; k \in [0,N-1]\}$. These furnish a projective representation of the symmetry in that $\mathsf g^e \mathsf g^o = \omega \mathsf g^o \mathsf g^e$ when acting on a single anyon.

This model also possesses a nontrivial anti-unitary symmetry,
which acts by complex conjugation.

Generalizations of the cluster state
(or {\it graph state})
to other groups, 
on bipartite graphs, 
are described here \cite{2015NJPh...17b3029B}.
However, only a stabilizer construction (and not the circuit construction that we require) is provided. 
It would be very interesting to generalize this construction to 
attach locally 1-dimensional SPTs to the string nets of arbitrary quantum double models.

\section{Stability of the physics of the snake monster}

\label{sec:snake-monster-stability}
\def\TT{{\bf T}}
Here we study the stability of the physics the cluster snake monster 
of \S\ref{sec:cluster-snake} with respect to
symmetric perturbations.
We show that the degenerate doublet 
is stable to small perturbations which preserve the $ \IZ_2 \times \IZ_2$ symmetry.
The basic claim is that the argument for stability \S\ref{sec:stability}
of the 1d SPT carries over to the 3d model.

Observe first that a naive hopping term for the electric defects $\TT_{ij} = \CZ_{ij}\sigma^x_{ij}$ is not symmetric under either $\IZ_2 \times \IZ_2$ or $\CT$. We can construct a symmetric hopping term if the anyon stays on the same sublattice: $\TT_{ac} = (1+Z^\nd_a Z^\nd_c)\CZ^\nd_{ab}\CZ^\nd_{bc} \sigma^x_{ab}\sigma^x_{bc}~$, where $b$ is a neighbor of both $a$ and $c$. 
This acts as a kinetic energy for the anyons and leads to identical dispersion relations $\epsilon(k) \sim \cos(k)$ for the two states comprising the doublet; they are degenerate at all momenta. 

\begin{wrapfigure}{r}{0.3\textwidth}
  \vspace{-30pt}
\begin{center}
\includegraphics[width=.3\textwidth]{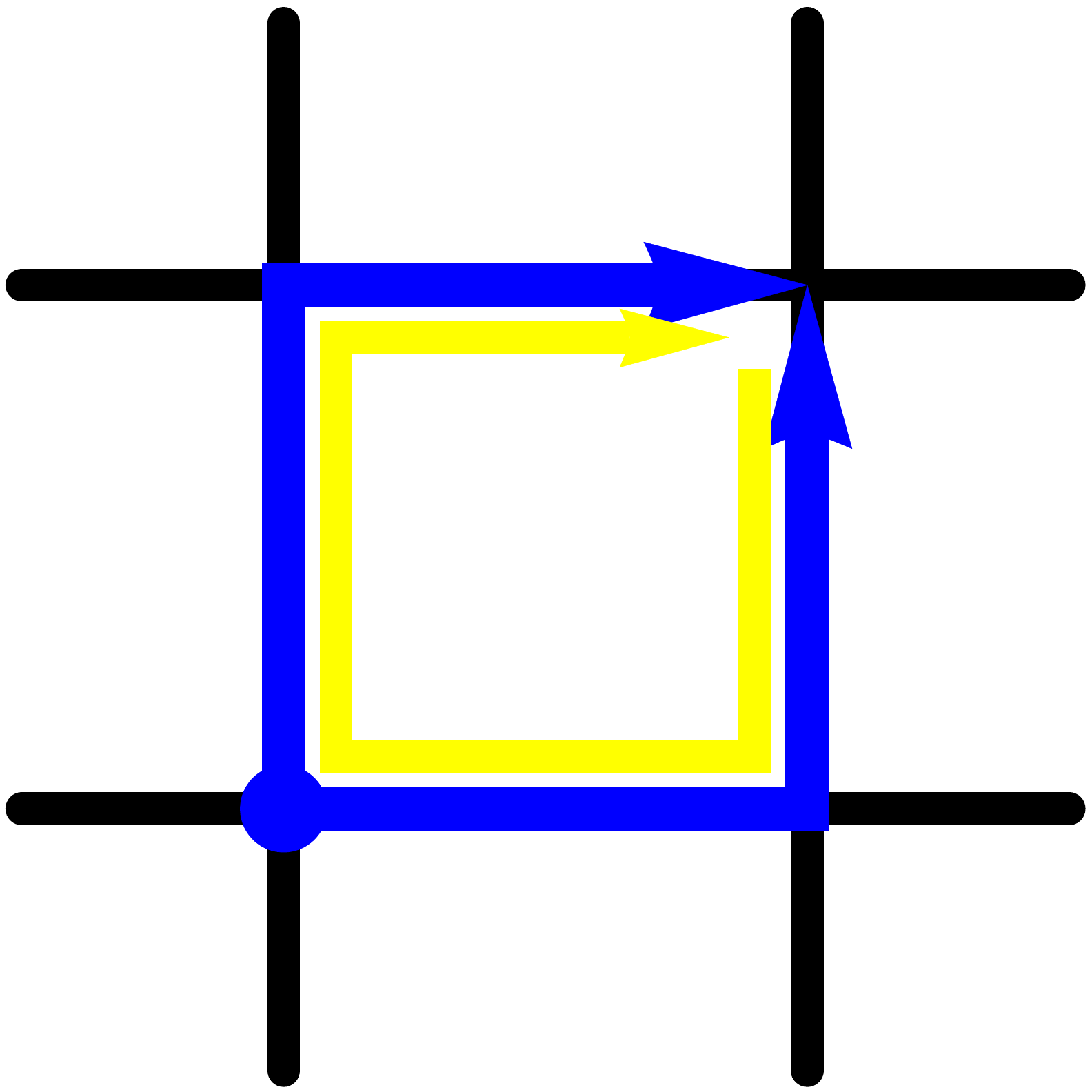}
%{fig-majorana-square-monster01.png}
\end{center}
  \vspace{-25pt}
%  \caption{Sod.\label{fig:sod}}
%  \vspace{-20pt}
\end{wrapfigure}
As an aside, we note that this realizes an Aharonov-Bohm effect. On the square lattice, there are two direct paths from one site to a same-sublattice-nearest-neighbor, illustrated to the right.
The difference between these two paths is a factor of $\CB_p$. Thus we see that the phase difference between the two paths is the flux through the plaquette.

Another possibility is to just add to the hamiltonian $\sum_l \sigma^x_l$, the kinetic term for the site defects in the `normal' toric code. In our case we need to check that the state obtained by acting with $\sigma^x_l$ on an anyon state actually produces another anyon state. To do so, we calculate the overlap of the states $\sigma^x_{bc}\ket{+/-}_b$ with the states $\ket{+/-}_c$. This calculation is illustrated for one of the four cases:
\be
\begin{split}
 \sigma^x_{bc}\ket{+} &= \sigma^x_{bc}W^{ab}\ket{\textrm{gs}} \\
 &= \sigma^x_{bc} \prod_{a\to b}\CZ_l \sigma^x_l \left(\frac{1}{\sqrt{\mathcal N_c}}\sum_C \prod_{l\in C}\CZ_l  \ket{\Rightarrow} \prod_{l\in C} \sigma^x_l \otimes_l \ket{\sigma^z_l = +1}\right)~.
\end{split}
\ee
We find the overlap of this state with an anyon state at the neighboring site $\ket{+}_c$
\be
\begin{split}
  &\frac{1}{\CN_c}\sum_{C,C'}  \left( \bra{\Rightarrow}\prod_{l'\in C'}\CZ_{l'} \prod_{l'\in C}  \otimes_{l'} \bra{\sigma^z_{l'} = +1}\sigma^x_{l'} \prod_{a\to c}\CZ^\nd_l \sigma^x_l\right) \sigma^x_{bc}\left( \prod_{a\to b}\CZ^\nd_l \sigma^x_l \prod_{l\in C}\CZ_l \ket{\Rightarrow} \prod_{l\in C} \sigma^x_l \otimes_l \ket{\sigma^z_l = +1}\right) \\
 &= \frac{1}{\CN_C}\sum_{C,C'} \delta_{C,C'}  \left(  \bra{\Rightarrow}\prod_{l'\in C'}\CZ_{l'}  \prod_{a\to c}\CZ^\nd_l \sigma^x_l\right) \sigma^x_{bc} \left(\prod_{a\to b}\CZ^\nd_l \sigma^x_l \prod_{l\in C}\CZ_l \ket{\Rightarrow} \right)\\
 &= \frac{1}{\CN_C} \sum_C  \bra{\Rightarrow} \CZ_{bc}  \ket{\Rightarrow} =\; _{bc}\bra{\rightarrow \rightarrow} \CZ_{bc} \ket{\rightarrow \rightarrow}_{bc}.
\end{split}
\ee
The overlap reduces to calculating the expectation value of the link operator in the ground state. The overlap of the other anyon states can be found by inserting factors of $Z_b$ or $Z_c$, yielding
\be
\label{eq:hopping_coefficients}
\begin{split}
 _c\bra{+}\sigma^x_{bc}\ket{+}_b &= 1/2 \\
  _c\bra{-}\sigma^x_{bc}\ket{+}_b &= 1/2 \\
 _c\bra{+}\sigma^x_{bc}\ket{-}_b &= 1/2 \\
 _c\bra{-}\sigma^x_{bc}\ket{-}_b &= -1/2 .
\end{split}
\ee

There are two other states produced by acting with $\sigma^x$ on an anyon state which have a defect in the site hamiltonian at the previous location of the anyon. In principle we should therefore assess the effect of this term through both degenerate and non-degenerate perturbation theory. However as we are interested in the dynamics of the anyons, degenerate perturbation theory leads to the relevant effect. Using the results given in \ref{eq:hopping_coefficients} we obtain the first order effective Hamiltonian
$$
\HH_{\textrm{eff}} \approx \sum_{ij,\sigma \sigma'} t_{\sigma \sigma'} c^\dagger_{i\sigma} c_{j\sigma'} + \textrm{h.c.}
$$
with
$$
t_{\sigma \sigma'} = \begin{pmatrix}
                      1 & 1\\
                      1 & -1
                     \end{pmatrix}.
$$

Diagonalizing this Hamiltonian leads to two hopping bands with energies proptional to $\pm \cos(ka)$. In this case the two bands are not identical; they are related by a $\pi$ phase shift. Every energy level at momentum $k$ has a degenerate mode in the other band at momentum $k+\pi$. The anyons still form doublets, albeit under an extended symmetry $\gG \times \gS$ where $\gS$ denotes the sublattice exchange operation. 
% \begin{figure}
% \centering
%  \includegraphics[width=0.5\textwidth]{band.eps}
%  \caption{Dispersion for the two spinons. The $\IZ_2$ symmetry acts to interchange the bands.}
% \end{figure}

\section{Self-dual models of confinement}

\label{app:pure-loop}

In this appendix we provide some context for the workings of the pure loop construction
of \cite{wns2015}, and generalize it to $\IZ_N$ strings.

The basic idea is to take a model of fluctuating string nets
and add an energetic penalty term which forbids nontrivial winding of the strings.
How do we impose a local condition which forces the strings to be contractible?

Here is a classical implementation which is well-known in certain corners of the statistical mechanics literature
(\eg~\cite{2000Chaikin-Lubensky}):
Consider a model with $\IZ_k$ variables $E_p = 0..k-1$ on the $d-1$-cells $p$ of a 
$d$-dimensional cell complex $\Delta$.
%\co
A configuration of such variables 
specifies by duality a string net (an assignment of 
$\IZ_k$ variables to the links of the dual cell complex $ \check \Delta$) $\check C$.
We will show that a sum over closed surfaces can produce the desired constraint
that this string net is contractible -- that is, 
it is the boundary of a collection of plaquettes. 

A sum over $ \IZ_N$ closed surfaces $S$
can be written as 
$$ \sum_{S, \text{closed}} ...
= \sum_{\{ \mu_p \} }  \prod_l \delta \( \sum_{q \in v(l)} \mu_q \) ...
= \sum_{\{ \mu_p \} }\sum_{\{ \alpha_l \}}
\omega^{\sum_l  \sum_{q \in v(l)} \mu_q \alpha_l }... 
$$
Consider the sum
$$ \sum_{S, \text{closed}} \omega^{ \oint_S \vec E \cdot d \vec a } 
= \sum_{\{ \mu_p \} }\sum_{\{ \alpha_l \}}
\omega^{\sum_l  \sum_{q \in v(l)} \mu_q \alpha_l }
\omega^{ \sum_p \mu_p E_p } $$
By definition of the vicinity operator, 
\be \label{eq:duality-sum} \sum_l \alpha_ l \sum_{ q \in v(l)} \mu_q =
 \sum_p \mu_p \sum_{l \in \partial p } \alpha_l \ee
Using this identity \eqref{eq:duality-sum}, we have 
$$ \sum_{S, \text{closed}} \omega^{ \oint_S \vec E \cdot d \vec a } 
= 
\sum_{\{ \alpha_l \}}
\sum_{\{ \mu_p \} }
\omega^{\sum_p \mu_p \( E_p  -  \sum_{l \in \partial p } \alpha_l\) } 
= \sum_{\{ \alpha_l \}} \prod_p \delta\( E_p -  \sum_{ l \in \partial p } \alpha_l \)  
 $$
This sum exactly imposes that $E_p$ is exact.

\subsection{$\IZ_N$ self-dual models of confinement}
\label{app:ZN-pure-loop}

\def\sX{\sigma^x}
\def\sZ{\sigma^z}

Consider two interpenetrating lattices $A$ and its dual lattice $B$ in three dimensions
(generalizations to other dimensions are interesting and will be discussed elsewhere).
Place $\IZ_N$ rotors on the links of the $A$ lattice (with operators $\sX_l, \sZ_l$),
and independent $\IZ_N$ rotors on the links of the $B$ lattice
(with operators $ \tau^{x,z}_p$);
these are in 1-to-1 correspondence with $d-1$-cells of the $A$ lattice
and we will label a link of $B$ by the $d-1$ cell of $A$ which it penetrates.
We try not to speak of cells of the $B$ lattice at all from now on.

%[Exercise: redo all of this for $k$-form variables on the $A$ lattice
%and $d-k$ form variables on the $B$ lattice.]

$$ H = - \sum_{l\in \Delta_1(A)} V( \CF_l^A )  - \sum_{p\in \Delta_{d-1}}  V( \CF_p^B )  $$
where $V$ is a real-valued function of a $\IZ_N$ variable with maximum when its argument is $1$, 
and 
$$ \CF_l^A \equiv \( \sZ_l\)^\dagger \prod_{p \in v(l)} \tau_p^x  ~~~~
\CF_p^B \equiv  \tau_p^z \prod_{l \in \partial p } \(\sX_l\)^\dagger ~~~.$$
Here 
all the links are counted with orientation, 
and $v$ is the vicinity operator, the oriented setwise inverse of the boundary map.

These operators all commute.
Their simultaneous unique eigenstate with eigenvalue $1$ has 
various useful representations:
\be \label{eq:MM2}\ket{\psi} =  \sum_{ \check\CM} \ket{\partial \check\CM_z}_A \otimes \ket{\check\CM_x}_{B} \ee
\be \label{eq:MM1} \ket{\psi} =  \sum_{ \CM} \ket{\CM_x}_A \otimes \ket{\partial \CM_z}_{B} \ee
\be\label{eq:CC} \ket{\psi} =  \sum_{C, \check C, \text{contractible}} \ket{C_z}_A \otimes \ket{\check C_z}_{B} \ee

More explicitly,  \eqref{eq:MM2} is 
\be \label{eq:MM21} \ket{\psi} =  
N^{- N_p/2} \sum_{\{ \mu_p \} } \ket{\mu_p }_B \otimes \ket{s_l =  \sum_{q \in v(l)} \mu_q}_{A} \ee
where $N_p$ is the number of links of the $B$ lattice, 
$ \ket{\mu_p}$ are $ \tau_p^x$ eigenstates and
$ \ket{s_l}$ are $\sZ_l$ eigenstates.
To see \eqref{eq:MM2}, apply the flip operators in the basis 
$$\ket{\psi} = \sum_{\{ \mu_p \} } \sum_{ \{s_l\} } 
\Phi(\mu, s) \ket{\mu_p }_B \otimes \ket{s_l}_A  ;$$
$ \CF_l^A  =1 $ requires $s$ to be a total divergence: $ s_l =  \sum_{q \in v(l) } \mu_q $, 
while $  \CF_p^B  =1 $  requires a uniform superposition of such states, 
by making the sheets hop.

Now let's discuss how to get from 
\eqref{eq:MM2} to \eqref{eq:CC}.
On a given link $p$, the $z$-basis and $x$-basis are related by
$$ \ket{\mu_p} = {1\over \sqrt{N}} \sum_{\sigma_p} \omega^{- \sigma_p \mu_p } \ket{ \sigma_p} $$
Therefore
\be \label{eq:MM22} \ket{\psi} =  
\sum_{{\sigma_p}}
\ket{\sigma}_B 
~ N^{- N_p} 
 \sum_{\{ \mu_p \} }
 \omega^{ -  \sum_p \sigma_p \mu_p } 
 \ket{s_l =  \sum_{q \in v(l)} \mu_q}_{A} \ee
Here 
$$ \sZ_l  \ket{ s, \sigma} = \omega^{s_l}  \ket{ s, \sigma}, ~~
\tau_p^z  \ket{ s, \sigma} = \omega^{\sigma_p}  \ket{ s, \sigma}. $$

We rewrite the sum over $\mu_p$ in two parts:
a membrane configuration $\mu$ on $\Delta_{d-1}(A)$ (the plaquettes $p$) can be decomposed as 
$$ \mu = \partial^{-1}(C) + S $$
where 
$S$ is a closed membrane, $ \partial S = 0 $.
$ C = \partial \mu$ is the boundary of the membrane $\mu$, 
a closed curve 
in $ \Omega_1(A) = \text{ker}(\partial) \subset \Delta_0(A)$.
$\partial^{-1}(C)$ is a {\it particular} fiducial membrane 
whose boundary is $C$.  
$S$ represents the deviation of $\mu$ from that choice.

Therefore
\be \label{eq:MM23} \ket{\psi} =  
\sum_{{\sigma_p}}
\ket{\sigma}_B 
~ N^{- N_p} 
\sum_{\{ \mu_p = \mu_p^0 + \hat \mu_p \} }
 \omega^{ -  \sum_p \sigma_p \mu_p^0 }
 \omega^{ -  \sum_p \sigma_p \hat \mu_p } 
 \ket{s_l =  \sum_{q \in v(l)} \mu_q^0}_{A} \ee
Here we have represented the plaquette sum as 
$$ \sum_{\{ \mu_p = \mu_p^0 + \hat \mu_p \} } .. 
=  \sum_{\{ \mu_p^0\} } \sum_{\{\hat \mu_p \} } ..
\equiv \sum_C \sum_S .. $$
and used 
the fact that the closed bit $\hat \mu_p$ satisfies by definition
$ \sum_{q \in v(l)} \hat \mu_q = 0 $
and therefore does not contribute to $s_l$.  Therefore:
\be \label{eq:MM23} \ket{\psi} =  
\sum_{{\sigma_p}}
\ket{\sigma}_B 
~ N^{- N_p} 
\sum_{C}
 \omega^{ -  \sum_p \sigma_p \mu_p^0 }
\ket{C}_A 
 \cdot 
\underbrace{
 \sum_{\{\hat \mu_p \} }
  \omega^{ -  \sum_p \sigma_p \hat \mu_p } 
}_{ \equiv \sum_S \omega^{ \oint_S \vec \sigma \cdot d\vec a } }
.
\ee
The underbraced factor 
is a (classical, \ie~no kets involved) 
sum over all closed $\IZ_N$ valued surfaces
weighted by the flux of a vector field through
those surfaces.  
%We do this sum next.
The result of this sum is to constrain $\{\sigma_p\} $ 
to only have support on contractible curves, $\check C$ 
\ie~ $\sigma_p = \sum_{l \in \partial p} \alpha_l $
for some set of link variables $ \alpha_l $.

The remaining factor from the fiducial membrane is then the linking number 
of these two sets of curves
$$ 
\omega^{ -  \sum_p \sigma_p \mu_p^0 } = 
\omega^{ l (C, \check C)} . $$
%These surfaces are sheets of magnetic flux
 
Using the classical formulae around \eqref{eq:duality-sum}, 
we can see explicitly that the fluctuating magnetic flux leads to confinement.
The sum over $ \hat \mu_p$ imposes that $\sigma_p$ is made of 
contractible curves $ \check C$ and 
we get \eqref{eq:CC}
$$ \ket{\Psi}  = 
\CN \sum_C \ket{C}_A 
\sum_{\check C}  \ket{\check C}_B ~\omega^{ l (C, \check C)} 
 $$ 
where $ \CN$ is a normalization factor.
 
Some comments:
\begin{enumerate}

\item Consider the $z$-basis representation (which will be \eqref{eq:CC}).
\be\label{eq:zzz} \ket{\psi} = \sum_{\{ s_l\}, \{ \sigma_p\}} \Psi(s, \sigma) \ket{ s, \sigma} ~.\ee
By combining $\CF$s we can make star operators on both sublattices:
$$ \prod_{ l \in v(j) } \CF^A_{l} = \prod_{ l \in v(j)} \(\sX_l\)^\dagger , \forall j \in \Delta_0(A)
~~~\prod_{ p \in \partial V } \CF^B_{p} = \prod_{ p \in \partial V} \tau_p^z,~
 \forall V \in \Delta_d(A).$$
 This means  that $ \Psi(s,\sigma)$ only has support on closed string configurations.

 \item Directly applying the flip operators to the $z$-basis representation \eqref{eq:zzz}
 we learn that 
\bea
\label{eq:FFF} \Psi(s, \sigma) &=& \omega^{\sigma_p} \Psi( s + \partial p, \sigma_p), ~~\forall p \in \Delta_{d-1}(A) 
\cr
 \Psi(s, \sigma) &=& \omega^{- s_l} \Psi( s, \sigma_p + v(l)), ~~\forall l \in \Delta_{1}(A) \eea
 
 \item By comparing to \eqref{eq:MM1} and \eqref{eq:MM2} we see that these closed
 strings must furthermore be contractible, since they are boundaries of membranes.
 This means that their linking number is well-defined.
 
\item The conditions \eqref{eq:FFF} are solved (up to normalization) by 
$$ \Psi(s, \sigma) = \omega^{l(s, \sigma)} $$
where $l$ is the linking number of the two configurations of closed surfaces.
A lattice formula for the linking number (from which we should be able to directly check
\eqref{eq:FFF}) is
$$ \omega^{l(s, \sigma)}  = \omega^{ \sum_p \sigma_p \sum \mu_p } |_{ s_l = \sum_{ p \in v(l) } \mu_p } $$
The expression for $\mu_p$ which solves 
$  s_l = \sum_{ p \in v(l) } \mu_p  $ 
is a lattice version of the 
Chern-Simons propagator.
\end{enumerate}

\vfill\eject

\bibliographystyle{ucsd}
\bibliography{collection} 
\end{document}